\documentclass{elsart}

\usepackage{epsfig,amssymb}

\begin{document}

\def \mucecay   {$ \mu^- \rightarrow \nu_\mu \, \bar{\nu}_e \, e^-$}
\def \amucecay  {$ \mu^+ \rightarrow \bar{\nu}_\mu \, \nu_e \, e^+$}
\def \numunutau {$ \nu_\mu \rightarrow \nu_\tau $}
\def \numunue   {$ \nu_\mu \rightarrow \nu_e $}
\def \numunus   {$ \nu_\mu \rightarrow \nu_s $}
\def \numu      {$ \nu_\mu $}
\def \nue       {$ \nu_e $}
\def \nutau     {$ \nu_\tau $}
\def \dm        {$\Delta m_{23}$}
\def \th        {$\theta_{23}$}

\begin{frontmatter}
\title{Neutrino cross-section measurement with neutrinos from muon decay}
\author{M.Campanelli},\ead{Mario.Campanelli@cern.ch}
\author{S.Navas-Concha},\ead{Sergio.Navas.Concha@cern.ch}
\author{A.Rubbia}\ead{Andre.Rubbia@cern.ch}
\address{Institut f$\ddot{u}$r Teilchenphysik, ETHZ, CH-8093 Z$\ddot{u}$rich, Switzerland}

\begin{abstract}
In this paper\footnote{Based on an invited talk given at the Third International Workshop on Neutrino Factories based on Muon Storage Rings (NUFACT01), May 2001, Tsukuba, Japan.}
 we stress the idea that new, more precise neutrino
cross-sections measurements at low energies will be necessary to improve
the results of future big neutrino detectors, which will be
dominated by the contribution of the systematic errors.
 The use of a muon beam instead of the traditional pion beams is proposed.
This choice allows the simultaneous measurement of both, \numu\ and
\nue\ interactions and the two helicities, in a clean environment
and with a precise knowledge of the beam flux.
 We show that with $\sim 10^{15} \; \mu$'s/year and a moderate mass detector
($\sim 100$ tons) placed close to the muon storage ring,
precisions of the order of 10\% in $\sigma_\nu / E_\nu$
($E_\nu$ bin size of 100~MeV) can be reached for neutrino energies below 2~GeV.
\end{abstract}

\end{frontmatter}
%
\vspace{-1.0cm}
\section{The neutrino cross-section measurements}
\label{sec:intro}
\vspace{-0.5cm}
 The motivations to measure the neutrino cross-sections at low energies
(below 5~GeV) are very strong.
For the current generation of large detectors with high accumulated statistics,
the {\it systematic errors } start to be a problem, dominating over
the statistical errors.
 At the present time, relatively few measurements of \numu\ cross-sections
at low energies exist (see Figure~\ref{fig:nuxiso-2gev}.left) showing, in general
poor precisions ($\Delta (\sigma_\nu / E_\nu) > $ 20\%)~\cite{BARKER}.
No measurements at all on \nue\ cross-sections have been published up to now.


 Several physics analysis could profit from a better knowledge of $\sigma_\nu \;$

\begin{itemize}
\vspace{-0.3cm}
\item[-] {\it Atmospheric neutrino analysis}.
 The present atmospheric neutrino measurements suffer from a poor knowledge
 of $\sigma_\nu \;$, which forces the use of double-ratios in order to cancel
 the systematics (the absolute rates are not reliable).
 The precision on the measurement
 of \dm\ and \th\ is limited by systematics on the flux and on the
 cross-sections. Also, the study whether the atmospheric neutrino anomaly
 is due to \numunutau\ or \numunus\ oscillations is limited by cross-sections
 uncertainties, and $\nu_\tau$ appearance search requires detailed
 understanding of final state topologies.
\item[-] {\it Proton decay searches}. A good knowledge of the $\nu$
 cross-section is required to evaluate in a reliable way the expected
 backgrounds. An accurate modeling of the background is needed for these
 channels that require background subtraction.
\item[-] {\it Long Baseline neutrinos}. One of the main goals of the Long
 Baseline neutrino programs is the precise measurement of the mixing parameters
 using appearance experiments, for which it is mandatory a careful
 understanding of the expected backgrounds.
\item[-] {\it Astrophysics}. The neutrino-nucleus cross-sections
 are also required for calculating certain astrophysical processes involving
 interactions in the outer shells of stars undergoing supernova explosions,
 which cause further nucleosynthesis to occur, changing the nuclear
 composition of the star.
\end{itemize}

 In addition, the nuclear targets vastly complicate the description
of the neutrino reactions. Effects like the target motion,
the binding energy or the final state interactions 
(specially important in proton decay searches)
can depend on the neutrino helicity and on the target type.
Even if there is a number of theoretical models trying to account for
nuclear effects in neutrino interactions, only a few comparisons
with experimental data are available.

 For all these reasons, it is clear that a new and precise measurement
of the neutrino cross-sections is needed.

\vspace{-0.5cm}
\section{Measurement of $\sigma_\nu$ : The idea and the assumptions}
\label{sec:idea}
\vspace{-0.5cm}
 The idea proposed here is based on the use of a low energy muon beam
as source of neutrinos.
This has two clear advantages compared to the traditional pion beams:
in one hand, it is possible to measure, at the same time, \numu\ and
\nue\ cross-sections and both helicities
(selecting the muon sign, \mucecay\ or \amucecay);
on the other, the measurement is done in a low contamination environment
and one profits from the precise knowledge of the muon flux.

 The analysis presented in this paper assumes a low energy
($E_{\mu} =$~2~GeV) and low intensity $\mu^-$ beam of $10^{15}$
{\it useful}\footnote{muons decaying on the direction of the detector}
muons/year. The detector is considered to be a cylinder of~1~m radius with
a moderate mass of~100~tons (a narrow and long detector clearly benefits
from the beam shape characteristics). 
Because of the limited geometrical acceptance of the detector, it is clear
that placing it as close as possible to the muon storage ring will increase
the flux of neutrinos traversing the cylinder: the shorter the baseline,
the higher the statistics and the smaller the error. Therefore, the baseline
is fixed to~10~meters. The length of the straight section of the muon storage
ring is fixed to~30~meters.

As shown in section~\ref{sec:analysis}, the above conditions allow
a measurement of $\sigma_\nu / E_\nu$ below 2~GeV with $\sim$10\% error
and a bin size of 100~MeV.
 Many channels can be studied under this assumptions
(and this counts for both, $\nu_\mu$ and $\nu_e$):

\begin{enumerate}
\item Charge  Current (CC) interactions ($\nu \; N \rightarrow l \; X$):
\begin{itemize}
\item[-] Quasi-elastic (QE) interaction ($\nu \; N \rightarrow l \; N'$)
\item[-] Single/Multi pion production ($\nu \; N \rightarrow l \; N' + n \, \pi$, $\; n > 0$)
\item[-] Resonant ($\nu \; N \rightarrow l \; N^{*} \;$ or $\; \nu \; N \rightarrow \nu \; N^{*}$)
\item[-] Deep-inelastic scattering ($\nu \; q \rightarrow l \; q' \;$ or
                                    $\; \nu \; q \rightarrow \nu \; q$)
\end{itemize}
\item Neutral Current (NC) interactions ($\nu \; N \rightarrow \nu \; N + n \,
\pi^0 \;$, $\; n \ge 0$)
\item Elastic electron scattering ($\nu \; e^- \rightarrow \nu \; e^-$):
\end{enumerate}

 The CC (NC) events are characterized by the presence (absence)
of a lepton in the final state. 
Specially important is the measurement
of the QE channels, since the $\nu_\mu$ QE interaction is the dominant
atmospheric reaction in a water \v{C}herenkov detectors like SuperK
($\sim$60\% of the total CC events).
A good particle identification and energy resolutions
are desired to separate between the charged ($\sim$27\%)
and the neutral ($\sim$10\%) single pion production channels.
The study of the Resonant interactions is specially relevant since it can be
an important source of background for proton decay events.
Finally, if the detector energy threshold for electrons is low enough
($<$10 MeV),
the measurement of the elastic electron scattering cross-sections can be
extremely useful for solar neutrino studies.
No particle miss-identification is assumed in this analysis.

\vspace{-0.5cm}
\section{ Results }
\label{sec:analysis}
\vspace{-0.5cm}

 As described in section~\ref{sec:idea}, we assume a flux
of intensity $10^{15}$ {\it useful} muons/year of~2~GeV energy.
The detector is considered to be a cylinder of~1~m radius with a
total mass of~100~tons. The baseline is~10~meters. Under these conditions,
the quoted numbers of expected $\nu_\mu$ and $\bar{\nu}_e$ events/year
from charge-current, quasi-elastic and resonant events can be found in
Table~\ref{tab:stat}.

\begin{center}
 \begin{figure}[t]
\begin{tabular}{cc}
 \epsfysize=7.0cm\epsfxsize=7.0cm
 \hspace*{-0.2cm}\epsffile{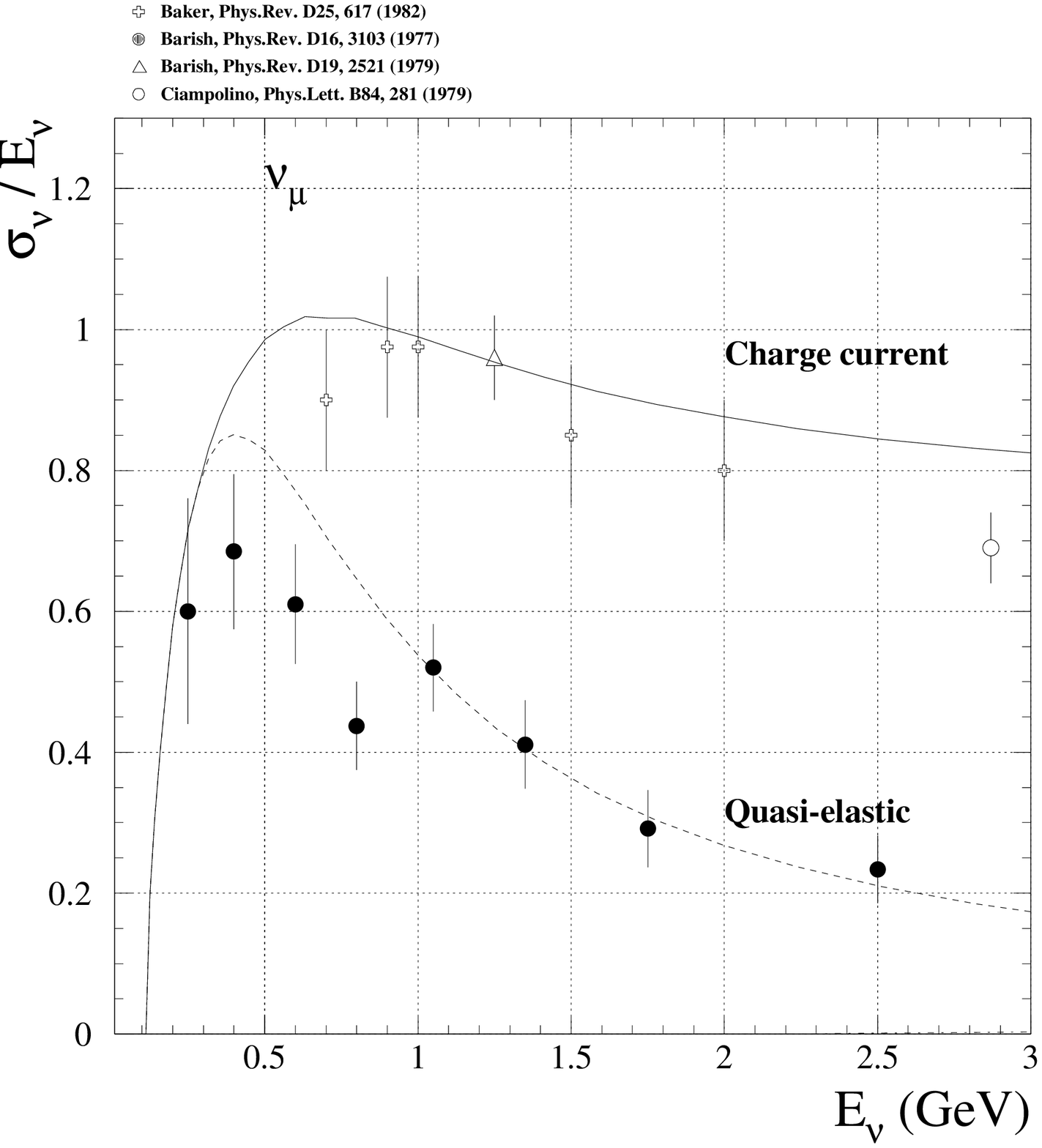} &
 \epsfysize=7.0cm\epsfxsize=7.0cm
 \hspace*{0.2cm}\epsffile{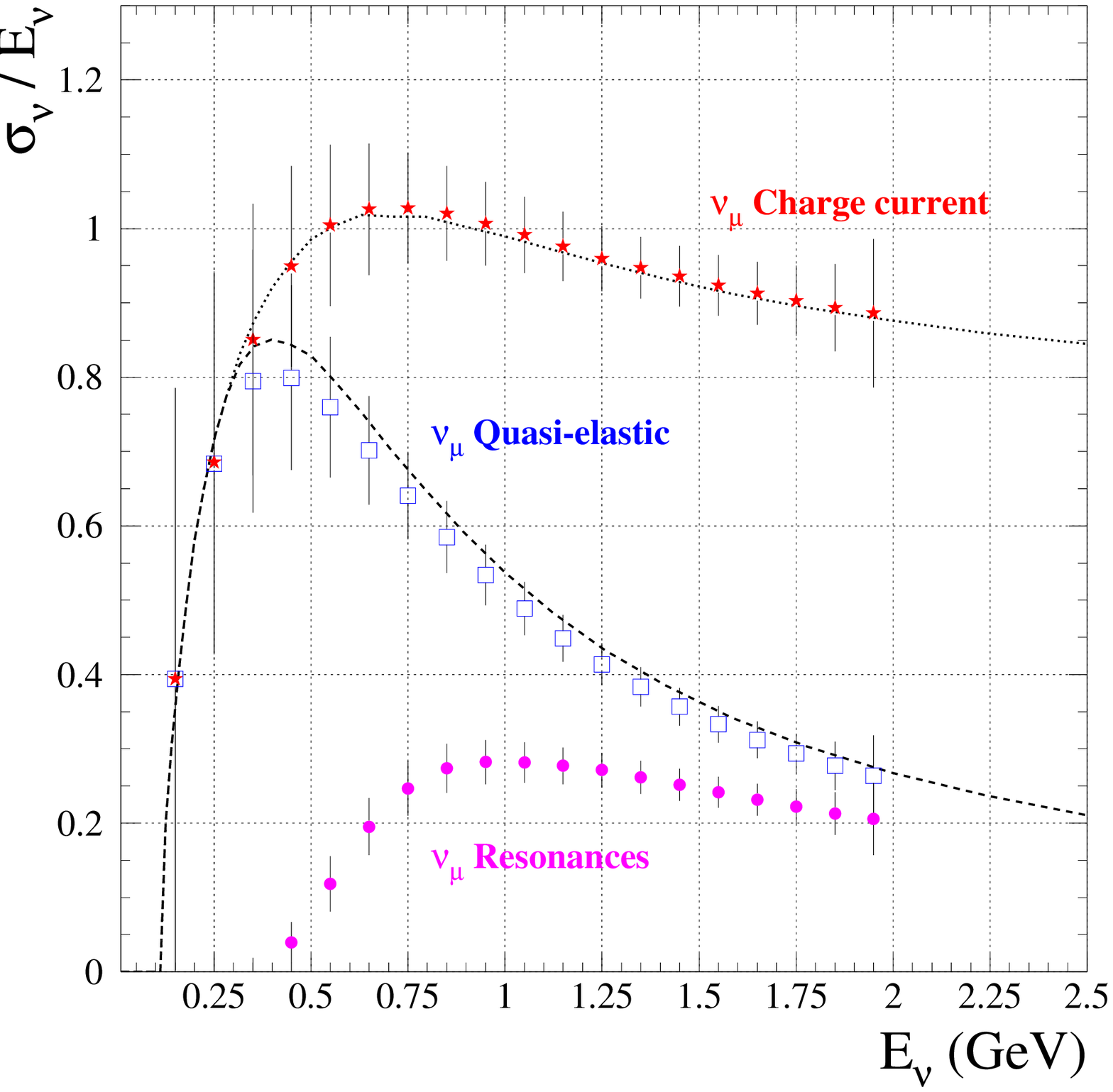}
\end{tabular}
\vspace{-0.0cm}
 \caption{\small Left: Survey of the present $\sigma_{\nu_{\mu}}$
  published data. Right: Obtained $\sigma_\nu / E_\nu$ spectra
 for $\nu_\mu$ interactions with the proposed method.}
 \label{fig:nuxiso-2gev}
\end{figure}
\end{center}

 The proposed set-up allows a clean measurement of the different neutrino
cross-sections with a small error and a high number of data points, as
shown in Figure~\ref{fig:nuxiso-2gev}.
 The four plots on the left part of the figure give the
relative errors on $\sigma_\nu / E_\nu$ as a function of the neutrino
energy ($\bar{\nu}_e$ and $\nu_\mu$) for charge-current and
quasi-elastic interactions. The bin size is 100~MeV. Above 0.5~GeV,
the largest part of the points exhibit an error smaller than 10\%.
Figure~\ref{fig:nuxiso-2gev} (right) shows the final obtained 
$\sigma_\nu / E_\nu$ spectrum for the three $\nu_\mu$ interactions.
The dotted lines are the theoretical predictions from the {\tt NUX}
model.

One attractive feature of the idea launched on this paper is that
with moderate means we can do a relevant measurement.
The requirements in terms of intensities and energies are far below
the neutrino factory needs, for instance
(a factor $\sim 10^{-5}$ a 10 less, respectively).
Also the required detector sizes (a factor $10^{-2}$ less) and the baseline
are much smaller.
Specially interesting should be to perform the cross-section
measurements using different types of detectors since some nuclear effects
(like final state interactions) depend on the nature of the target
(water, Iron, Argon, etc.). Moreover, this can allow a test of the full
detector (measuring efficiencies, backgrounds, etc.) in real conditions and
much before the starting of the neutrino factory program.

\begin{table}[h]
\begin{center}
\begin{tabular}{|c|c|c|} \cline{2-3}
 \multicolumn{1}{}{ } &
 \multicolumn{1}{|c|}{$\nu_\mu$} &
 \multicolumn{1}{|c|}{$\bar{\nu}_e$}    \\
\hline
Total Charge-Current (QE+DIS)   & 5600 & 1590  \\ \hline
Quasi-Elastic interactions (QE) & 2460 &  740  \\ \hline
Resonances                      & 1450 &  600  \\ \hline
\end{tabular}
\caption{\small
Expected number of CC, QE and Resonant events/year
from $\nu_\mu$ and $\bar{\nu}_e$ interactions.}
\label{tab:stat}
\end{center}
\end{table}

\end{document}